\documentclass[12pt]{iopart}
\usepackage[utf8]{inputenc}
\usepackage{graphicx}
\usepackage{bm}
\usepackage[dvipsnames]{xcolor}
\usepackage[normalem]{ulem}
\usepackage{textcomp}
\usepackage{lineno,hyperref}
\usepackage{hyperref}
\usepackage{xspace}
\usepackage{siunitx}
\usepackage{times}
\usepackage{amssymb}
\usepackage{svg}

\usepackage{upgreek}
\usepackage{pdfpages}
\newcommand{\WSe}{WSe$_2$\xspace}
\newcommand{\SiO}{SiO$_2$\xspace}
\newcommand{\WTe}{WTe$_2$\xspace}

\newcommand{\NbSe}{NbSe$_2$\xspace}
\newcommand{\rmnum}[1]{\romannumeral #1}

\begin{document}

\title[\footnotesize Suspended pick-up and flip-over assembly of van der Waals heterostructures with ultra-clean surfaces]{Suspended dry pick-up and flip-over assembly for van der Waals heterostructures with ultra-clean surfaces}

\author{K Jin$^{1,2,3}$, T Wichmann$^{1,2,4}$, S Wenzel$^{1,2}$ , T Samuely$^5$, O Onufriienko$^5$, P Szabo$^5$, K Watanabe$^6$, T Taniguchi$^7$, J Yan$^8$,  F S Tautz$^{1,2,4}$, F L\"upke$^{1,2}$, M Ternes$^{1,2,3}$, J Martinez-Castro$^{1,2,3,*}$}

\address{$^1$Peter Gr\"unberg Institut (PGI-3), Forschungszentrum J\"ulich, 52425 J\"ulich, Germany}

\address{$^2$J\"ulich Aachen Research Alliance, Fundamentals of Future Information Technology, 52425 J\"ulich, Germany}
\address{$^3$Institute for Experimental Physics II B, RWTH Aachen, 52074 Aachen, Germany}

\address{$^4$Institute for Experimental Physics IV A, RWTH Aachen, 52074 Aachen, Germany}

\address{$^5$Centre of Low Temperature Physics, Faculty of Science, Pavol Jozef Šafárik University \& Institute of Experimental Physics, Slovak Academy of Sciences, 04001 Košice, Slovakia}
\address{$^6$Research Center for Electronic and Optical Materials, National Institute for Materials Science, 1-1 Namiki, Tsukuba 305-0044, Japan}
\address{$^7$Research Center for Materials Nanoarchitectonics, National Institute for Materials Science,  1-1 Namiki, Tsukuba 305-0044, Japan}
\address{$^8$Materials Science and Technology Division, Oak Ridge National Laboratory, Oak Ridge, TN 37831, USA}

\ead{j.martinez@fz-juelich.de}

\begin{abstract}
Van der Waals heterostructures are an excellent platform for studying intriguing interface phenomena, such as moir\'e and proximity effects. 
Surface science techniques like scanning tunneling microscopy (STM) have proven a powerful tool to study such heterostructures but have so far been hampered because of their high sensitivity to surface contamination. 
Here, we report a dry polymer-based assembly technique to fabricate van der Waals heterostructures with atomically clean surfaces.
The key features of our {\it suspended dry pick-up and flip-over technique} are
1) the heterostructure surface never comes into contact with polymers, 
2) it is entirely solvent-free,  
3) it is entirely performed in a glovebox, and
4) it only requires temperatures below \SI{130}{\celsius}.
By performing ambient atomic force microscopy and atomically-resolved scanning tunneling microscopy on example heterostructures, we demonstrate that we can fabricate air-sensitive heterostructures with ultra-clean interfaces and surfaces.
Due to the lack of polymer melting, the technique is further compatible with heterostructure assembly under ultra-high vacuum conditions, which promises ultimate heterostructure quality.

\end{abstract}
\maketitle

\section{Introduction}
The mechanical assembly of van der Waals (vdW) heterostructures \cite{castellanos-gomez2022} is a  key technology for studying the emerging phenomena occurring at interfaces between 2D materials \cite{Cao2018,Wang2022}. The popularity of this method is based on the ease and speed with which heterostructures can be built in a virtually infinite number of possible combinations. The constant demand for cleaner and better devices has driven the successive refinement of the assembly techniques, with the aim to meet the criteria of reliability, cleanliness, and interface quality to an ever greater extent, despite the simultaneous increase in the complexity of the structures. Currently, the \textit{dry pick-up assembly} of 2D materials \cite{Castellanos-gomez2014} and related techniques \cite{Pizzocchero2016,Son2020,Wakafuji2020, rebollo2021} are the most widely used ones in this regard as they offer great versatility and yield high quality heterostructures. 
The method of dry pick-up assembly proceeds as follows: A polydimethylsiloxane (PDMS) polymer stamp covered with a sticky polymer film is used to pick up 2D crystals exfoliated from bulk material onto a substrate, typically Si/SiO$_2$. While the polymer film has a strong adhesion over a certain  temperature range, the PDMS is soft, allowing a controlled contact to the flakes and preventing them from damage. The heterostructure is then assembled from the top to the bottom, starting with  the material that will eventually form the topmost layer. In the last step, the heterostructure is released by contacting the target substrate and melting the polymer. Alternatively, a polymer whose thermoplastic properties allows a melt-free release can be employed \cite{Wakafuji2020}. 
Crucially, standard dry pick-up assembly techniques require fully contacting the heterostructure surface with the polymers, which inevitably leaves polymer residues on the surface of the final heterostructure.
Thus, if the heterostructure surface is formed by a reactive material, the contact with the polymer will degrade its surface.

A strategy to circumvent this problem is the use of a protective layer, i.\,e., an inert material that is placed as the topmost layer of the heterostructure to protect (encapsulate) the layers underneath from degradation
\cite{Son2020,Cao2015Quality,Qiu2021,Xu2022}. 
In this case, the polymer residues can be removed from the protective layer with a combination of solvents such as chloroform, acetone or isopropanol, and with AFM-based mechanical cleaning methods \cite{Goossens2012}.
The combination of inert encapsulation layers with such cleaning methods has been demonstrated to allow  contamination-sensitive surface science techniques to be applied successfully \cite{Martinez2018Scanning,Cucchi2019}. 
Nevertheless, protective layers can pose a limit to the accessibility of the underlying material due to their finite thickness and electronic properties which can mask those of the underlying materials. 

An alternative to avoid polymers to come into contact with the heterostructure surface is to assemble it in reverse order and to flip it over after the assembly process \cite{Kyounghwan2016, lupke2020,Luepke2022}.
Using such techniques, the vdW heterostructure is typically released by melting the polymer layer (polypropylene carbonate, PPC) that is used during assembly, leaving a thick PPC layer underneath the finished heterostructure. 
The PPC is then removed by annealing at \SI{250}{\celsius} in vacuum.
Unfortunately, while the dry-transfer flip technique is a huge step forward, it has its limitations: 
1) The assembled heterostructure must be annealed in high vacuum for several hours to properly remove the residual PPC layer from underneath the heterostructure. Such extended annealing can cause atomic defects on the heterostructure surface \cite{lupke2020}.
2) The required annealing temperature of \SI{250}{\celsius} is incompatible with many 2D materials which degrade at this temperature \cite{Shcherbakov2018}. 
3) The large amount of PPC that must be evaporated makes it incompatible with ultra-high vacuum (UHV) environments, which are sensitive to organic contaminants. 

Here, we present a novel technique for the mechanical assembly of ultra-clean vdW heterostructures which overcomes the drawbacks of existing assembly methods. Our \textit{suspended dry pick-up and flip-over} technique does not require the use of a protective encapsulation layer, nor does it require additional cleaning such as by solvents or AFM.
We demonstrate that we achieve atomically clean surfaces for a variety of assembled heterostructures using ambient AFM and low-temperature STM characterization.



\section{Methods}
\subsection{Stamp preparation\label{sec: method}}

\begin{figure}[tbh]
	\centering
 \includegraphics{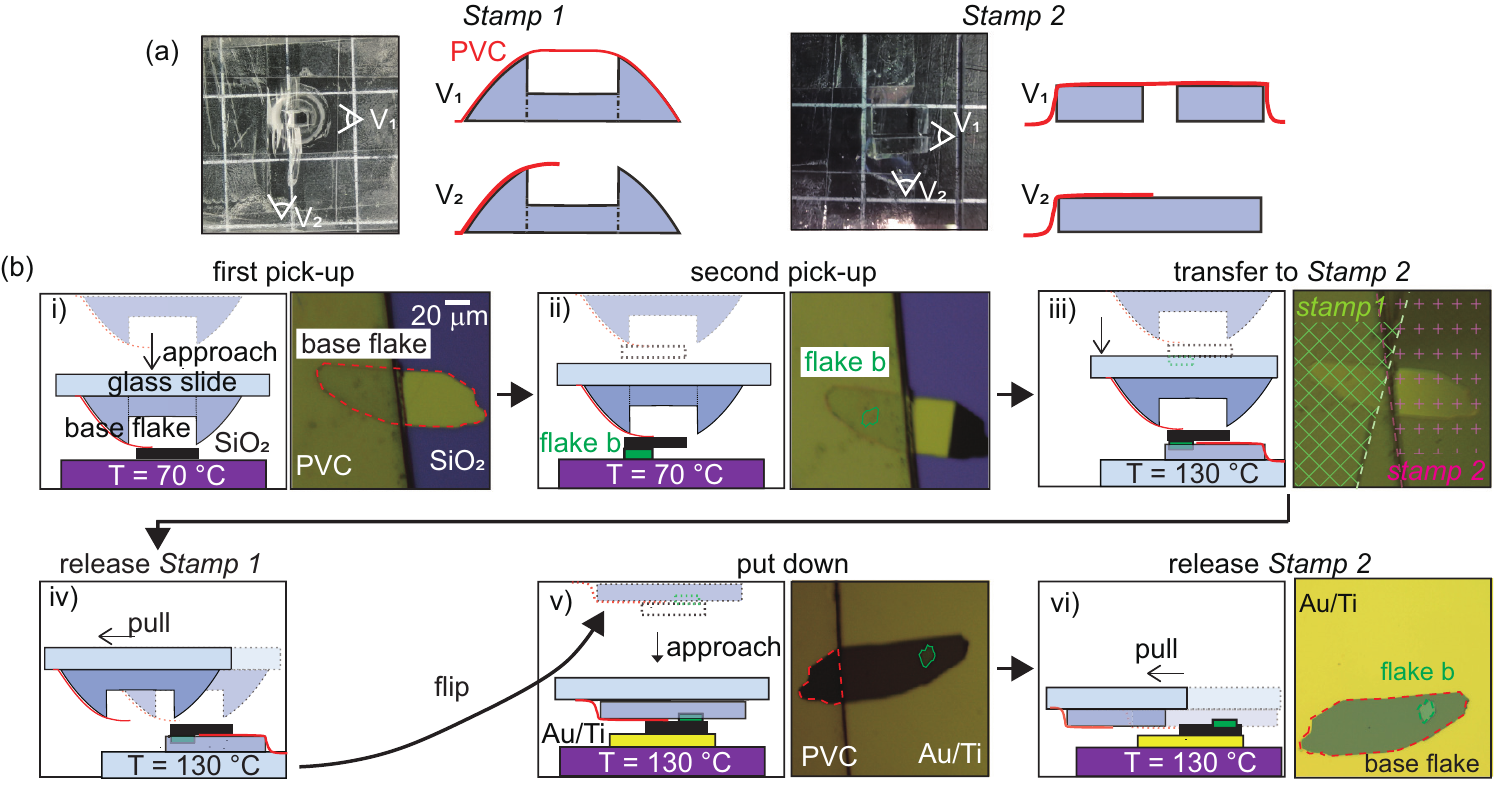}
	\caption{\label{fig1}
	{\bf Schematic representation of the \textit{suspended dry pick-up and flip-over} assembly of a \WSe/Graphite heterostructure}. \textbf{(a)} Top view optical micrographs of polymer stamp 1 and stamp 2	and side view schematics (from indicated perspectives $V_1$ and $V_2$), detailing how the PVC films (red) are suspended from the PDMS (grey). 
	\textbf{(b)} Flow diagram of the heterostructure assembly. 
	\textbf{(i)} Stamp 1 is brought into contact with the base flake (here graphite) at \SI{70}{\celsius} by touching the left half of the base flake (black) with the PVC film. The base flake is then picked up by carefully retracting stamp 1 vertically. 
	\textbf{(ii)} The van der Waals heterostructure is assembled by contacting the exfoliated flake b (green, \WSe),
	initially located on the SiO$_2$  substrate (purple), with the part of the base flake that is located directly underneath the PVC film. 
	\textbf{(iii)} The base flake on stamp 1 is brought into contact with stamp 2 at \SI{130}{\celsius}, making sure that each stamp touches roughly half of the base flake (white and red dashed lines). 
	\textbf{(iv)} The van der Waals heterostructure is transferred by sliding stamp 1 sideways until the base flake is only in contact with stamp 2. 
	\textbf{(v)} After flipping the heterostructure upside down, stamp 2 with the heterostructure is mounted in the micromanipulator stage. Subsequently, Stamp 2 is brought into contact with the target substrate Au/Ti (yellow) on SiO$_2$ (purple) at \SI{130}{\celsius}. 
	\textbf{(vi)} The heterostructure is finally released from stamp 2  by sliding the latter sideways and retracting it vertically.
	}
\end{figure}

We start by preparing two PDMS stamps. First, we used a commercial PDMS elastomer kit (SYLGARD\textregistered 184), mixing a polymeric base and curing agent at a 10:1 (w/w) ratio in a Petri dish. To prepare the dome-shaped PDMS for stamp 1, the Petri dish was placed upside down for several days, allowing the mixed liquid to slowly form a droplet and cure simultaneously.) The first one (stamp 1) is used to assemble the vdW heterostructure, while the second one (stamp 2) is used to flip and deterministically place the vdW heterostructure onto the target substrate. Stamp 1 uses a dome-shaped PDMS, with a $500\times \SI{500}{\micro\meter\squared}$ square depression cut into its center, and is supported on a microscope glass slide (Fig.\,\ref{fig1} \textbf{(a)}). Stamp 2 consists of a flat, \SI{5}{\milli\meter} thick PDMS block that has been cut into two pieces with a scalpel, resulting in a trench of approximately \SI{200}{\micro\meter} width (Fig.\,\ref{fig1} \textbf{(b)}) that exposes the supporting glass slide. For more details on the transfer slide preparation, see Supp.\,Fig.\,S1.
  
Next, we prepare the polymer films that will be used for pick-up, flip-over, and release of the vdW heterostructure. To this end, a commercial  poly(vinyl chloride) (PVC) film (RIKEN WRAP, Riken Fabro Corp)  \cite{Wakafuji2020} is first annealed to \SI{130}{\celsius} for one minute on a hot plate. This step is crucial because it prevents any uncontrolled thermal shrinkage of the PVC when mounted on the stamps and heated during subsequent steps (Supp.\,Fig.\,S2). After annealing, we transfer the PVC film onto a $1 \times \SI{1}{\centi\meter\squared}$ square of standard double-sided scotch tape, taking care not to wrinkle the film. The double-sided tape provides support and stability when manipulating the PVC film. Next, we cut the PVC film into two pieces and place one of them on PDMS stamp 1, the other on PDMS stamp 2 (see Supp.\,Fig.\,S3), covering half of the square depression and half of the trench as shown in Fig.\,\ref{fig1}a and b. Note that, both, the detailed shape of the PDMS stamps as well as the total contact area of the PVC films with the PDMS determine the stiffness of the suspended PVC films. 
In detail, we have observed that stiffer films show stronger adhesion with 2D materials.
Thus, by placing the PVC on the two stamps such that the suspended area on stamp 1 is larger than on stamp 2, we promote stronger adhesion between polymer and heterostructure on stamp 2 compared to stamp 1.
This crucial feature enables us to transfer the assembled vdW heterostructure from stamp 1 to stamp 2 later in the process, despite them being chemically identical PVC films.

\subsection{Sample preparation}
Having prepared the stamps, we exfoliate flakes from bulk crystals  onto \SI{285}{\nano\meter} SiO$_2$/Si substrates. Before the exfoliation, the SiO$_2$/Si substrates are ultrasonically cleaned by acetone and 2-propanol, followed by \SI{30}{\second} UV-Ozone treatment\cite{huang2015}. The assembled vdW hetero\-structures were placed either onto a \SI{285}{\nano\meter} SiO$_2$/Si substrate or alternatively onto pre-evaporated $\SI{100}{\nano\meter}/\SI{10}{\nano\meter}$ Au/Ti leads on a \SI{285}{\nano\meter} SiO$_2$/Si substrate and mounted to a standard STM sample plate for the STM measurements. All samples have been fabricated in an argon-filled glovebox.

\subsection{Heterostructure assembly}
We assembled the vdW heterostructure in reverse order using a standard assembly stage (HQ graphene) as described in Figure \ref{fig1} \textbf{(b)}.
We note that the first flake to be picked up, the base flake, should be thicker than $\approx$ \SI{40}{\nano\meter} to provide the necessary stiffness and mechanical support for the subsequently following layers of the heterostructure. We contact the base flake with stamp 1 at \SI{70}{\celsius} by touching roughly half of the flake with the PVC film and then carefully pulling the stam  p up vertically (panel (\rmnum{1}) in Fig.\,\ref{fig1}b and Supp.\ Video 1). Subsequently, we pick up subsequent flakes by vdW interaction \cite{Pizzocchero2016} at \SI{70}{\celsius} to assemble the desired heterostructure.
For each additional pick-up step, we make sure that the flake does not extend over the edge of the base flake ( panel (\rmnum{2}) in  Fig.\,\ref{fig1}b and Supp.\,Video 2).

Once we have completed the vdW heterostructure assembly on stamp 1, we flip the heterostructure using stamp 2. 
To do this, we place stamp 2 on the transfer stage and raise the temperature to \SI{130}{\celsius}. We then carefully bring stamp 1 into contact with stamp 2, making sure that the PVC films on the two stamps only touch the base flake on opposite sides and not each other (panel (\rmnum{3}) in Fig.\,\ref{fig1}b). 
Stamp 1 is then released by gently pressing and sliding it laterally (panel (\rmnum{4}) in Fig.\,\ref{fig1}b and Supp.\,Video\,3). 
Next, stamp 2 is flipped over by mounting it into the micro-manipulator.
The heterostructure is then brought into contact with the target substrate at \SI{130}{\celsius} (panel (\rmnum{5}) in Fig.\,\ref{fig1}b). Finally, we release the vdW heterostructure by slowly moving stamp 2 laterally ( panel (\rmnum{6}) in Fig.\,\ref{fig1}b and Supp.\,Video\,4). 

\subsection{Atomic force microscopy}
Atomic force microscopy (AFM) experiments were conducted using a Bruker Innova instrument under ambient condition. Contact-mode AFM was used with a setpoint force of $\sim$\SI{6.2}{\nano\newton} and a scan speed of \SI{20}{\micro\meter\per\second}. The probing tips used in the experiments were of type Bruker RESPA-20, with a nominal tip radius of \SI{8}{\nano\meter} and a spring constant of \SI{0.9}{\newton\per\meter}.

\subsection{Scanning tunnelling microscopy}
Scanning tunneling data were acquired at the Centre of Low Temperature Physics in Košice in ultra-high vacuum at a base pressure of $\sim$\SI{1e-10}{\milli\bar} and a base temperature of \SI{1.14}{\kelvin} using a mechanically cut Au tip.

\section{Results and discussion}
\subsection{Example heterostructures}

\begin{figure}[tbh]
	\centering
	\includegraphics{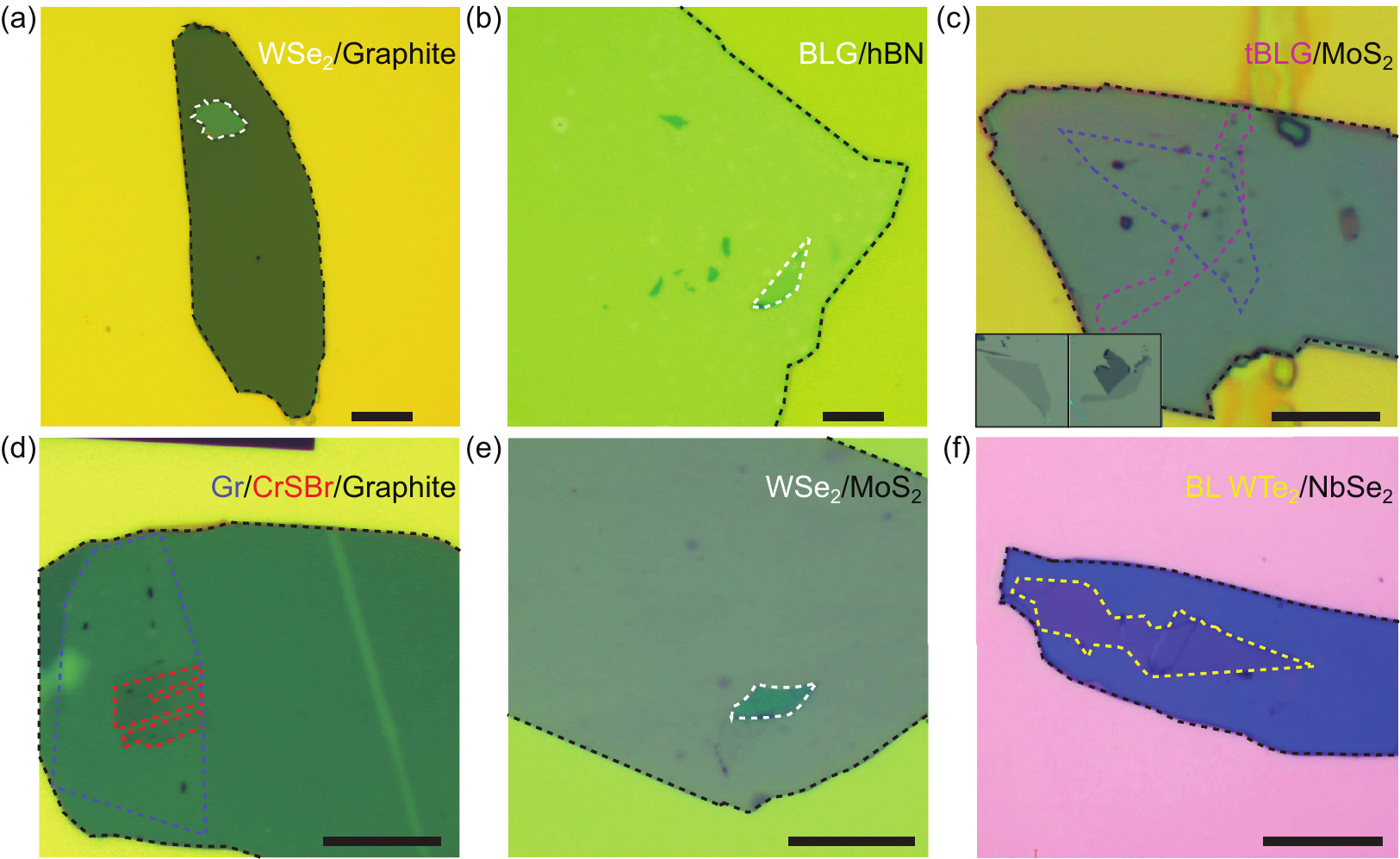}

	\caption{\label{fig:fig2}
	{\bf Exemplary vdW heterostructures.}
	All panels show optical microscopy images of various assembled heterostructures on Au covered \SiO substrates.
	The flake outlines are indicated as color-coded dashed lines.
	{\bf (a)} WSe$_2$ on  graphite. {\bf (b)}  Bilayer graphene on  hexagonal boron nitride. {\bf (c)} Twisted-bilayer graphene on MoS$_2$. Inset: optical images of the exfoliated graphene flakes prior to the assembly. {\bf (d)} Graphene on CrSBr on graphite. {\bf (e)} WSe$_2$ on MoS$_2$. {\bf (f)} BL WTe$_2$ on NbSe$_2$. The scale bars are \SI{20}{\micro\meter} in all panels.} 
\end{figure}

By assembling example heterostructures composed of a wide variety of different vdW materials we demonstrate the broad applicability of our suspended assembly technique.
The used base flakes of the heterostructures include common materials typically used for optical, transport, and surface characterization experiments, such as hexagonal boron nitride ($h$-BN), graphite, or MoS$_2$ \cite{Icking2022,Sigger2022,Parashar2023}. 
In addition, we demonstrate that air-sensitive (reactive) materials, such as NbSe$_2$, can also be used as base flakes, although only the base flake surface area which did not come into contact with the PVC film is atomically clean. 
Figure\ \ref{fig:fig2} shows a summary of six example heterostructures:
({\bf a}) WSe$_2$ on graphite, the former being a transition metal dichalcogenide known to induce strong spin-orbit coupling into graphene \cite{Fulop2021};
({\bf b}) bilayer graphene on $h$-BN, the latter being an insulator which is typically used for gating purposes \cite{Sui2015};
({\bf c}) twisted-bilayer graphene on MoS$_2$, assembled from two separate graphene monolayers that have been sequentially  picked up with a controlled twist angle between them, allowing the simultaneous study of the strongly correlated physics of twisted-bilayer graphene and the proximity-induced strong spin-orbit coupling by the transition metal dichalcogenide \cite{Wang2016}; 
({\bf d}) graphene on CrSBr, the latter being a 2D antiferromagnetic semiconductor whose electronic interlayer coupling can be magnetically controlled \cite{Wilson2021};
({\bf e}) WSe$_2$ on MoS$_2$, a semiconductor heterostructure that has been intensively studied for its optoelectronic properties \cite{Kim2017}; ({\bf f}) WTe$_2$ on NbSe$_2$, a heterostructure that realizes one-dimensional topological superconductivity \cite{lupke2020, Martinez-Castro2023}. 

From the successful assembly of the heterostructures demonstrated here, we conclude that the \textit{suspended dry pick-up and flip-over} assembly technique can be applied to a wide range of materials including most of the common materials used for optics, electronic transport, and surface science studies.
We note that, if the are any additional principal limitations at all, they will most likely be found in the vdW heterostructure assembly process itself: A too strong interaction between the exfoliated flakes and the \SiO substrate might prevent the pick-up by the base flake or previously picked-up layers on top of the base flake.

\subsection{Interface quality}
\begin{figure}[tbh]
	\centering
	\includegraphics[scale=1]{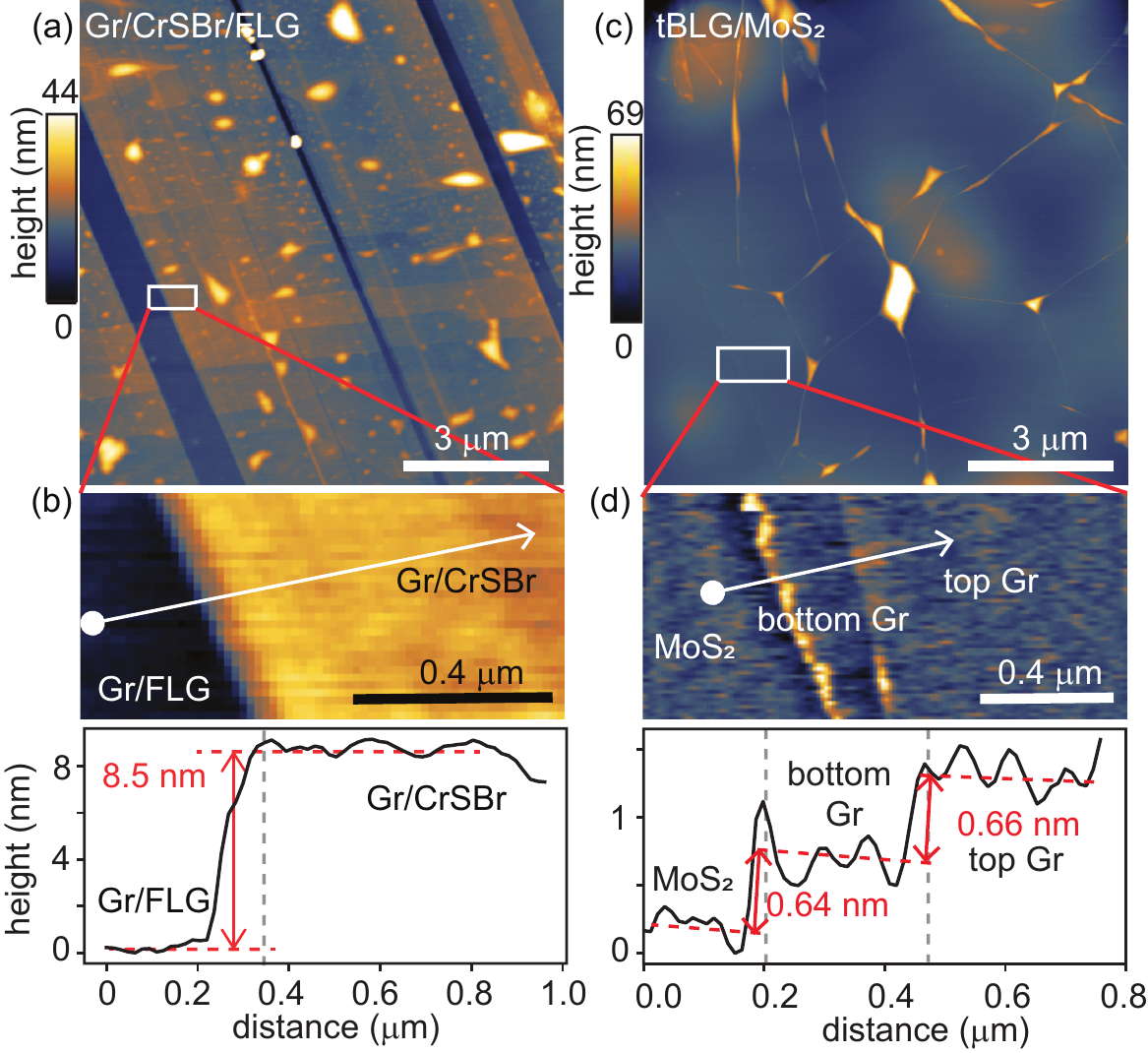}
	\caption{\label{fig:fig3}
	{\bf Atomic force microscopy on assembled vdW heterostructures}
    {\bf (a)} AFM topography of graphene on CrSBr on few-layer graphite (FLG). 
    {\bf(b)} Zoom into the region highlighted by the white rectangle in panel a and the topography cross section along the white arrow. The thicknesses of the CrSBr steps are indicated.  
    {\bf(c)} AFM topography of twisted-bilayer graphene on MoS$_2$. {\bf(d)} Zoom into the region highlighted by the white rectangle in panel (\textbf{c}), showing MoS$_2$ and the two graphene layers and topography cross section along the white arrow. The thicknesses of the steps correspond to that of the graphene layers.} 
\end{figure}

To assess the quality of our assembled vdW heterostructures, we characterize the resulting surface and internal interface quality in the following. 
To this end, we inspect the assembled vdW heterostructures with ambient condition contact-mode AFM (c-AFM), focusing here on those with top layers of either monolayer graphene (Fig.\,\ref{fig:fig3} (\textbf{a}), (\textbf{b})) or twisted bilayer graphene (Fig.\,\ref{fig:fig3} \textbf({c}), \textbf({d})), because compared to the thicker 2D materials, graphene is more prone to wrinkle and fold. 
Less than optimal transfer methods tend to produce more trapped blisters and wrinkles in the graphene, allowing a direct comparison between different mechanical assembly methods. 
We assess three different criteria: 
1) The number and area of trapped blisters at the interface, 
2) possible damage of the top-most layer such as ruptures, 
3) the amount of residue left at the surface.
In comparison to standard dry-transfer techniques, we find that a reverse assembly of vdW heterostructures generally results in a lower number of trapped blisters and protects the top-most surface from rupturing  in agreement with earlier reports \cite{Onodera2022}.
Furthermore, the overall higher stiffness of the heterostructure provided by the base flake contributes to better interfaces, because it avoids inhomogeneities and stress arising from the polymer film flexing. 
Lastly, we do not observe any traces of residue or damage on the topmost graphene layer whatsoever. 
In fact, the only visible residues are located in the contact front between the PVC and the base flake (see Supp.\,Fig.\,S4). 

To further compare the surface quality achieved with our suspended assembly technique to standard dry-transfer methods, we assembled a heterostructure following Ref.~\cite{Wakafuji2020}, where the heterostructure surface comes in full contact with the PVC polymer, i.e. without flip. 
The surface of the resulting Gr/NbSe$_2$/graphite heterostructure  (Supp.\,Fig.\,S5) exhibits substantial polymer residue as well as damaged regions where the graphene is partially ruptured. 
In contrast, the vdW heterostructures assembled with our suspended assembly technique do not show as damage or surface contamination.
Detailed analysis of the AFM topographies in Fig.~\ref{fig:fig3} (\textbf{b}), (\textbf{d}) shows that their roughness is smaller than the noise of the c-AFM (RMS = \SI{0.74}{\nano\meter}), demonstrating the flatness of the vdW heterostructures' interfaces assembled with this method.

\subsection{Scanning tunneling microscopy}

\begin{figure}[tbh]
	\centering
	\includegraphics[scale=1]{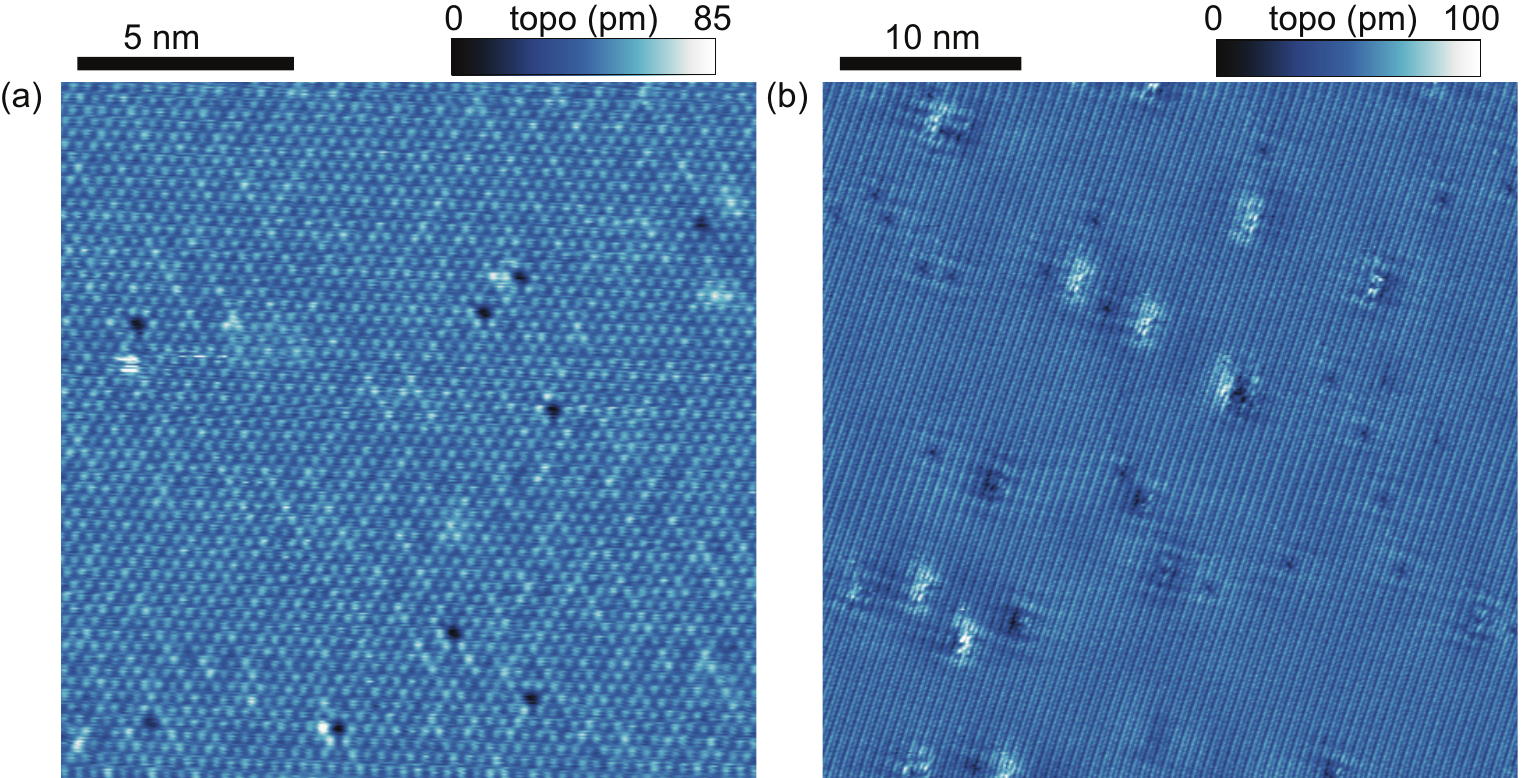}
	\caption{\label{fig:fig5}
	{\bf Scanning tunneling microscopy on a BL \WTe/\NbSe heterostructure} 
	{\bf (a)} Atomic resolution STM topography on bulk NbSe$_2$ showing its $3\times 3$ charge density wave (sample bias $V_\text{bias}=\SI{90}{\milli\volt}$ and tunneling current  $I_\text{set}=\SI{100}{\pico\ampere}$). {\bf (b)} Atomic resolution STM topography on bilayer \WTe on bulk NbSe$_2$, $V_\text{bias}=\SI{100}{\milli\volt}$, $I_\text{set}=\SI{100}{\pico\ampere}$). Only a few single-atomic defects are observed on both surfaces, i.e. close to the material's bulk defect concentrations.
} 
\end{figure}

Finally, we analyzed the surface quality and thus the compatibility of our suspended assembly technique with contamination-sensitive surface science techniques by STM. For this, we inspected the \WTe/\NbSe vdW heterostructure shown in Fig.~\ref{fig:fig2} (\textbf{f}).
Since both \NbSe and \WTe are air-sensitive, the heterostructure was assembled in an argon-filled glovebox and transferred from there to the UHV chamber hosting the STM in a vacuum suitcase. 
Fig.~\ref{fig:fig5} (\textbf{a}) shows an atomically resolved image of NbSe$_2$---without any surface contaminants---and its $3\times 3$ charge density wave, which emerges below $T_\text{CDW}=\SI{32}{\kelvin}$ and is known to be very sensitive to external perturbations \cite{Giambattista1998}. 
In the same way, we were able to obtain atomically resolved images of the bilayer \WTe situated on the NbSe$_2$, again showing only single atomic defects.

\section{Conclusion}
The developed transfer technique allows the assembly of vdW heterostructures with ultra-clean surfaces and interfaces. Notably, the method is suitable for heterostructure assembly in a glovebox and thus can be used for air-sensitive materials. The high quality of the resulting vdW heterostructures enables detailed surface studies, such as  STM with atomic resolution even on air-sensitive materials, without requiring a protective layer. 
Our \textit{suspended dry pick-up and flip-over} assembly technique can be readily applied to the \textit{zoo} of available vdW materials and therefore can be applied in various research areas such as optics, electronic transport, and surface science. 
Because the  does not require polymer melting nor chemical solvents, it also constitutes an important step towards the all-UHV fabrication of vdW heterostructures.


\par

\section{Acknowledgements}
The authors thank Fran\c{c}ois C. Bocquet for technical support.  Furthermore, we are grateful to the Helmholtz Nano Facility for its support regarding sample fabrication. The authors acknowledge funding from the European Union’s Horizon 2020 Research and Innovation Programme under Grant Agreement no 824109 (European Microkelvin Platform). J.M.C., T.W., K.J., M.T. and F.L. acknowledge funding by the Deutsche Forschungsgemeinschaft (DFG, German Research Foundation) within the Priority Programme SPP 2244 (project nos. 443416235 and 422707584). J.M.C., F.S.T. and F.L. acknowledge funding from  the Bavarian Ministry of Economic Affairs, Regional Development and Energy within Bavaria’s High-Tech Agenda Project ''Bausteine f\"ur das Quantencomputing auf Basis topologischer Materialien mit experimentellen und theoretischen Ans\"atzen''.  J.M.C. acknowledges funding from the Alexander von Humboldt Foundation. T.S., P.S. and O.O. acknowledge the support of APVV-20-0425, VEGA 2/0058/20, Slovak Academy of Sciences project IMPULZ IM-2021-42, COST action CA21144 (SUPERQUMAP) and EU ERDF (European regional development fund) Grant No. VA SR ITMS2014+ 313011W856. S.W. and F.S.T. acknowledge funding by the DFG through the SFB 1083 Structure and Dynamics of Internal Interfaces (project A12). M.T. acknowledges support from the Heisenberg Program (Grant No. TE 833/2-1) of the German Research Foundation. F.L. acknowledges financial support by Germany’s Excellence Strategy - Cluster of Excellence Matter and Light for Quantum Computing (ML4Q) through an Independence Grant. J.Q.Y. was supported by the US Department of Energy, Office of Science, Basic Energy Sciences, Materials Sciences and Engineering Division. K.W. and T.T. acknowledge support from the JSPS KAKENHI (Grant Numbers 20H00354, 21H05233 and 23H02052) and World Premier International Research Center Initiative (WPI), MEXT, Japan.

\section*{Data availability statement}
All data of this study are available at the JÜLICH DATA repository under the doi .... 

\section*{References}

\bibliography{ref.bib}

\newpage

\setcounter{figure}{0}
\setcounter{section}{0}
\renewcommand{\thefigure}{S\arabic{figure}}
\title{\textnormal{Supplemental information}
	\\Suspended dry pick-up and flip-over assembly of van der Waals heterostructures with ultra-clean surfaces }

\author{K Jin$^{1,2,3}$, T Wichmann$^{1,2,4}$, S Wenzel$^{1,2}$ , T Samuely$^5$, O Onufriienko$^5$, P Szabo$^5$, K Watanabe$^6$, T Taniguchi$^7$, J Yan$^8$,  F S Tautz$^{1,2,4}$, F L\"upke$^{1,2}$, M Ternes$^{1,2,3}$, J Martinez-Castro$^{1,2,3,*}$}

\address{$^1$Peter Gr\"unberg Institut (PGI-3), Forschungszentrum J\"ulich, 52425 J\"ulich, Germany}

\address{$^2$J\"ulich Aachen Research Alliance, Fundamentals of Future Information Technology, 52425 J\"ulich, Germany}
\address{$^3$Institute for Experimental Physics II B, RWTH Aachen, 52074 Aachen, Germany}

\address{$^4$Institute for Experimental Physics IV A, RWTH Aachen, 52074 Aachen, Germany}

\address{$^5$Centre of Low Temperature Physics, Faculty of Science, Pavol Jozef Šafárik University \& Institute of Experimental Physics, Slovak Academy of Sciences, 04001 Košice, Slovakia}
\address{$^6$Research Center for Electronic and Optical Materials, National Institute for Materials Science, 1-1 Namiki, Tsukuba 305-0044, Japan}
\address{$^7$Research Center for Materials Nanoarchitectonics, National Institute for Materials Science,  1-1 Namiki, Tsukuba 305-0044, Japan}
\address{$^8$Materials Science and Technology Division, Oak Ridge National Laboratory, Oak Ridge, TN 37831, USA}

\ead{j.martinez@fz-juelich.de}


\maketitle

\section{Stamp preparation}

\begin{figure}[htbp]
	\centering
	\includegraphics[scale=1]{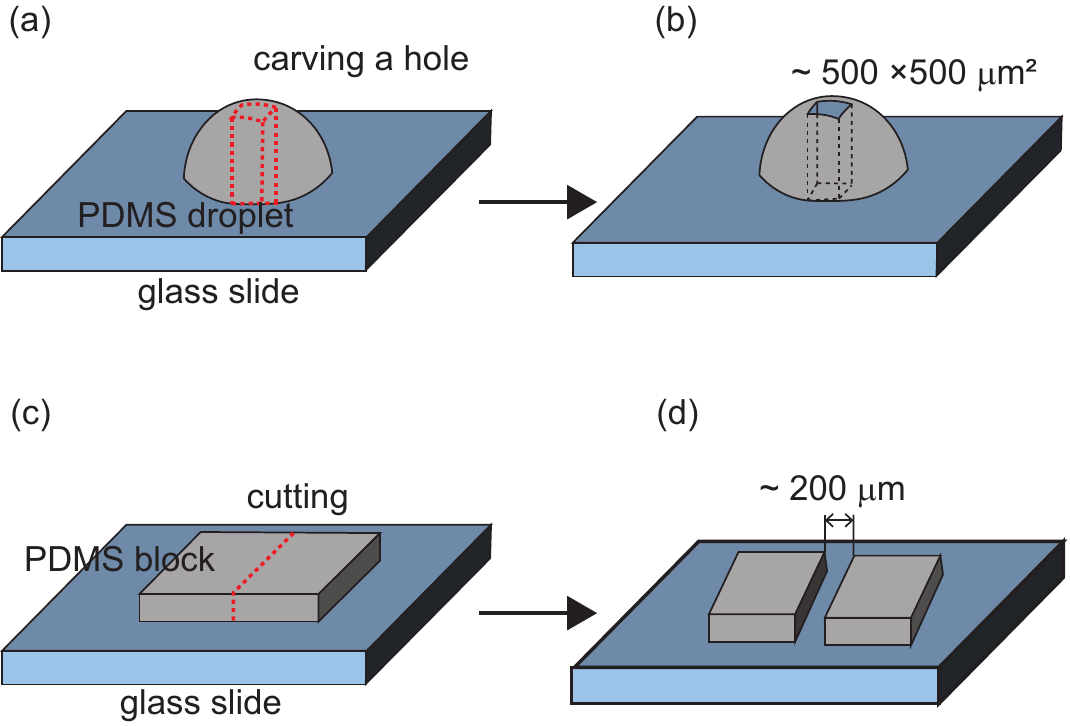}
	\caption{{\bf Preparation of the PDMS supporting structure for stamp 1 and 2.} {\bf(a)} The red dotted lines indicate the volume of the dome-shape PDMS to be carved out in {\bf (b)}. {\bf(c)} The red dotted marks where the PDMS will be split in two {\bf (d)}, to form a trench of $\approx$\SI{200}{\micro\meter}. 
		\label{fig:S2}}
\end{figure}

The fabrication of the transfer stamps involves two steps: the preparation of the supporting PDMS and the PVC film.
Supp. fig. \ref{fig:S2} provides a schematic of the supporting PDMS preparation. Stamp 1 starts with a PDMS dome shape whose center has been carved out with the help of a scalpel and, for simplicity, in the shape of a square (Supp. Fig. \ref{fig:S2} (\textbf{a})). 
Stamp 2 consists of a rectangular block of PDMS of $\approx$\SI{5}{\milli\meter} thickness (Supp. Fig. \ref{fig:S2}(\textbf{c})). We then used a scalpel to carefully cut the PDMS into two rectangles, separating them and forming a trench of approximately \SI{200}{\micro\meter} (Supp. Fig. \ref{fig:S2} (\textbf{d})). To achieve good optical visibility for the vdW heterostructure assembly, we made a trench with sharp edges by cutting the PDMS as perpendicular to the surface as possible. \par

\newpage
\section{PVC preparation process}

\begin{figure}[htbp]
	\centering
	\includegraphics[scale=1]{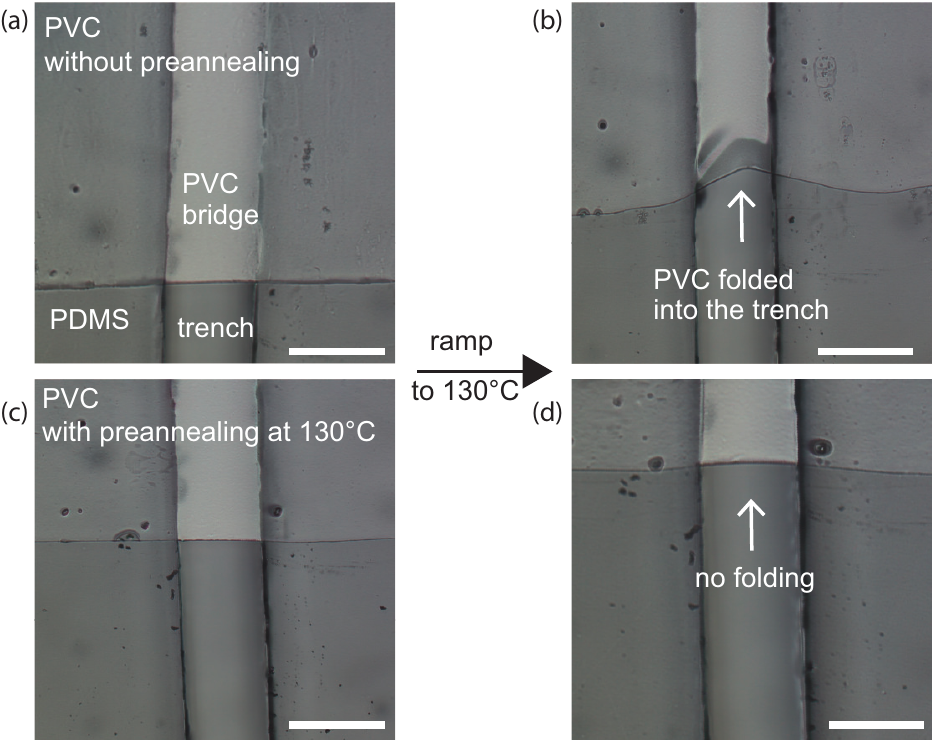}
	\caption{{\bf Analysis of the suspended PVC mechanical behaviour with and without annealing at \SI{130}{\celsius}.} {\bf(a)} Optical image of the PVC film mounted on stamp 2 without pre-annealing treatment. {\bf (b)} Optical image of the same stamp in (a) after being heated to \SI{130}{\celsius}. {\bf(c)} Optical image of a stamp as described in (a) on which the PVC has been pre-annealed to \SI{130}{\celsius} prior to bringing it into contact with the supporting PDMS. {\bf(d)} Optical image of the stamp in (c) after being annealed at \SI{130}{\celsius} for one minute. The scale bar is \SI{200}{\micro\meter} for all the panels. 
		\label{fig:S1}
	}
\end{figure}

The success of vdW heterostructures transfer between stamps relies on the mechanical stability when the suspended PVC film is heated to \SI{130}{\celsius}. A non pre-annealed PVC film will retract and wrinkle when mounted on stamp 2 and simply impede the transfer. Supp. Fig. \ref{fig:S1} shows the difference between a PVC film subjected to pre-annealing and a PVC film mounted directly on the stamp without pre-annealing. Specifically, a PVC film mounted on the stamp without being pre-annealed (Supp. Fig. \ref{fig:S1} (\textbf{a})) will severely fold towards the trench when heated at a temperature of  \SI{130}{\celsius}  (Supp. Fig. \ref{fig:S1}(\textbf{b})). This behaviour is avoided when the PVC film is pre-annealed, keeping PVC film shape and stability (Supp. Fig. \ref{fig:S1} (\textbf{c}),(\textbf{d})).  

\newpage
\section{Mounting PVC films on the supporting PDMS}

\begin{figure}[htbp]
	\centering
	\includegraphics[scale=1]{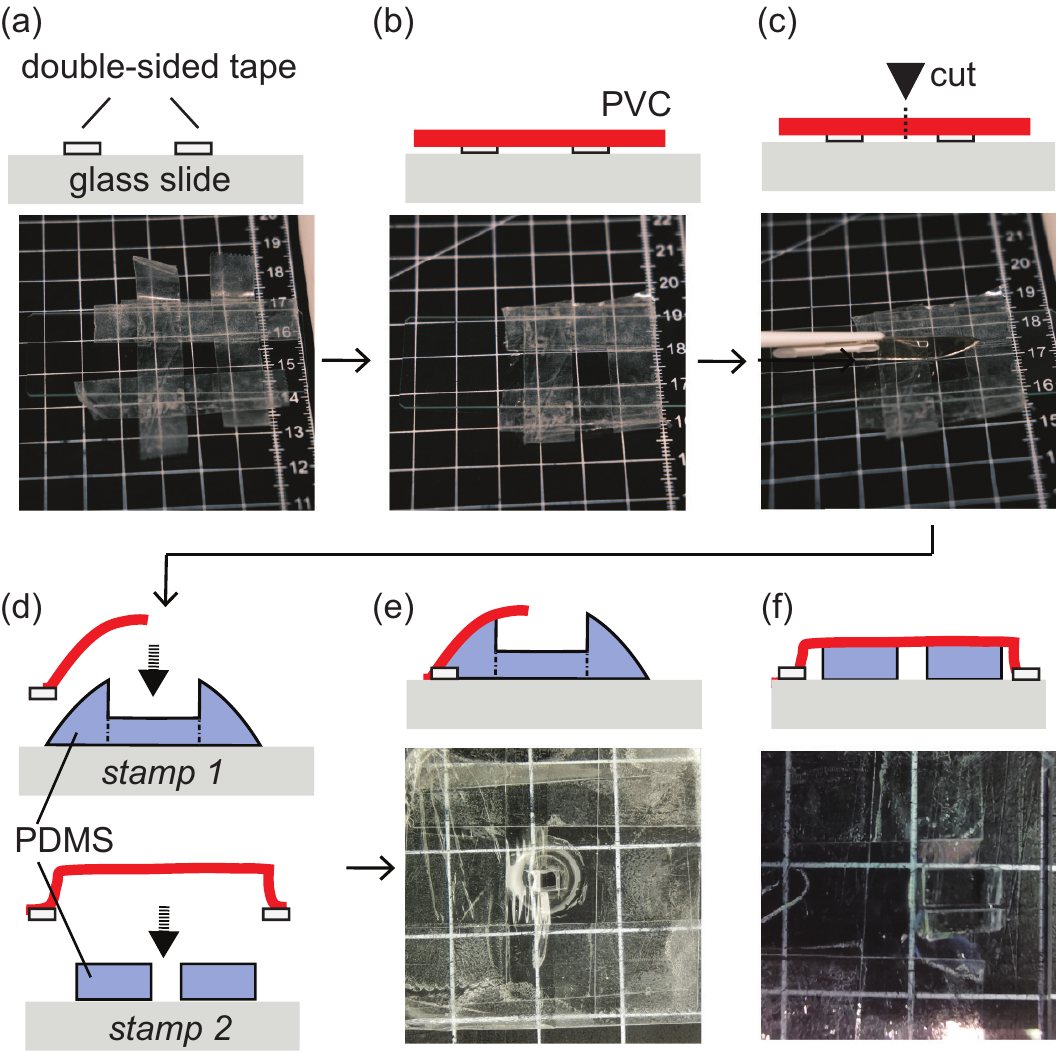}
	\caption{{\bf Stamp preparation.} {\bf (a)} A square made out of double-sided scotch tape is prepared on top of the clean surface of a glass slide. {\bf (b)} The pre-annealed PVC film is transferred on top of the double-sided scotch tape. {\bf (c)} The PVC film adhered to the double-sided scotch tape is cut in two with a sharp scalpel. {\bf (d)}  The PVC films with the supporting double-sided frame are transferred to the supporting PDMS places on glass slides. {\bf (e,f)} The PVC films are fixed to the stamps partially suspended on the carved-out square in the dome shape PDMS and the trench. 
		\label{fig:S3}
	}
\end{figure}

\newpage

\section{Boundary between the PVC-contact and suspend region}

\begin{figure}[htbp]
	\centering
	\includegraphics[scale=1]{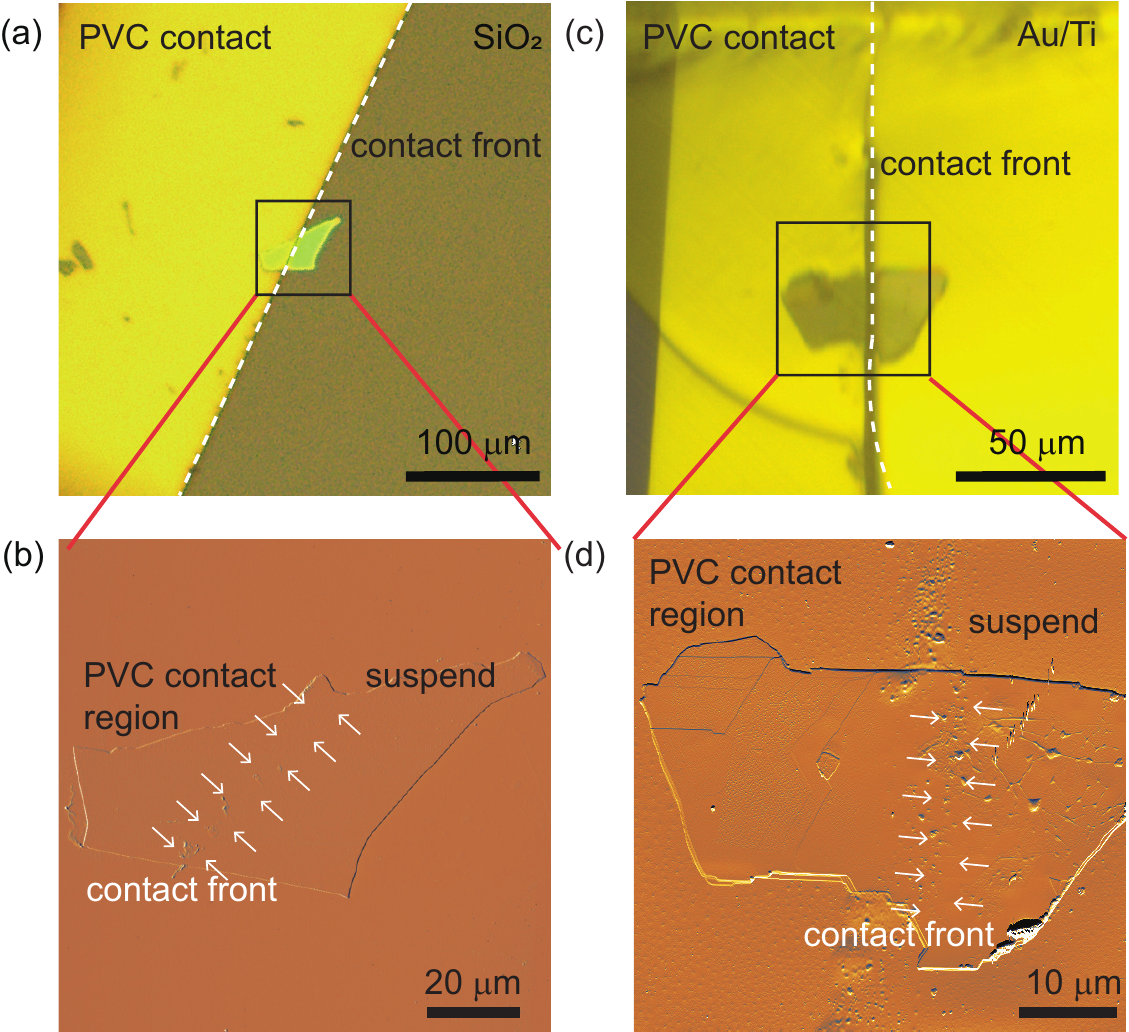}
	\caption{{\bf Determination of polymer residue with contact atomic force microscopy.} {\bf(a)}  Optical image of a partially suspended graphite flake prior to release. The white dashed line indicates the contact front between the PVC film and the graphite flake. {\bf(b)} AFM error-signal image of the graphite flake after transferring it to a \SiO chip.  {\bf(c)} Optical image of the partially suspended tBLG/MoS$_2$ heterostructure. {\bf(d)} AFM error-signal image of the tBLG/MoS$_2$ heterostructure after transferring it to a \SiO chip.  White arrows in (b) and (d) indicate the location of the polymer residue coinciding with the PVC film contact. \label{fig:S5} }
\end{figure}

\begin{figure}[htbp]
	\centering
	\includegraphics[scale=1]{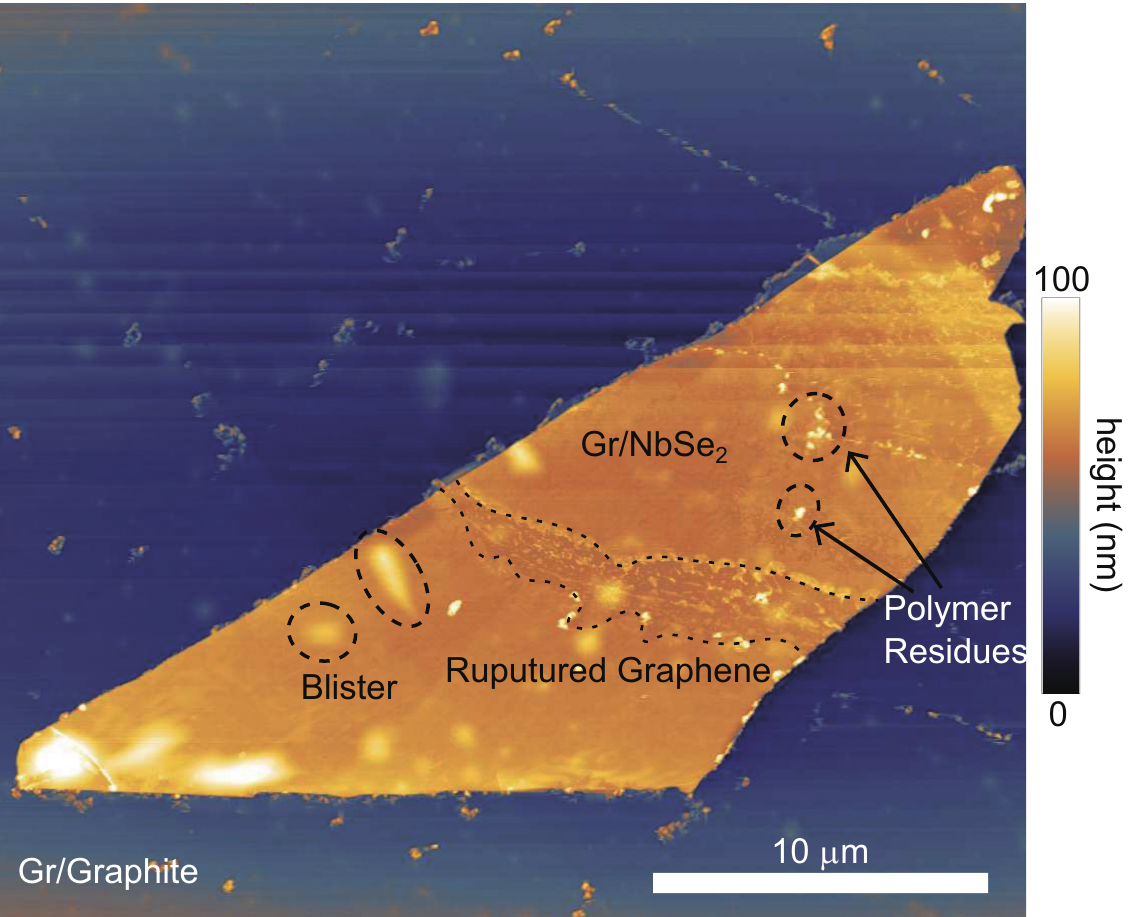}
	\caption{{\bf AFM image of the graphene-encapsulated NbSe$_2$ bulk on graphite prepared by the PVC/micro-dome PDMS full contact transfer.} The black dashed circles and lines mark the prominent defects of this heterostructure, including graphene cracks, blisters, and polymer residue. \label{fig:S4}}
\end{figure}

\end{document}